\documentclass[conference]{IEEEtran}

\usepackage{acro}

\acsetup{single}

\usepackage{balance}
\usepackage{cite}

\AtBeginDocument{\bstctlcite{IEEEexample:BSTcontrol}}

\usepackage{etoolbox}
\usepackage[T1]{fontenc}
\usepackage{graphicx}
\usepackage{mathtools}
\usepackage{microtype}
\usepackage{mleftright}

\AtBeginDocument{\mleftright}

\usepackage{newtx}

\DeclareEncodingSubset{TS1}{ntxtlf}{0}

\usepackage{siunitx}

\usepackage{soul}

\makeatletter
\NewCommandCopy{\oldmakecaption}{\@makecaption}
\RequirePackage[font = footnotesize, labelformat = simple]{subcaption}
\RenewCommandCopy{\@makecaption}{\oldmakecaption}
\makeatother

\usepackage{tabularray}

\UseTblrLibrary{booktabs}

\AtEndPreamble{
  \usepackage[capitalize]{cleveref}

  \crefname{equation}{}{}
  \Crefname{equation}{Eq.}{Eqs.}
  \Crefname{figure}{Fig.}{Figs.}
}

\DeclareAcronym{FD-SOI}{short = FD-SOI, short-indefinite = an, long = fully depleted silicon on insulator}
\DeclareAcronym{IP3}{short = IP3, short-indefinite = an, long = third-order intercept point}
\DeclareAcronym{IIP3}{short = IIP3, short-indefinite = an, long = input \ac{IP3}}
\DeclareAcronym{IM3}{short = IM3, short-indefinite = an, long = third-order intermodulation}
\DeclareAcronym{LO}{short = LO, short-indefinite = an, long = local oscillator}
\DeclareAcronym{RF}{short = RF, short-indefinite = an, long = radio frequency}
\DeclareAcronym{RF-DAC}{short = RF-DAC, short-indefinite = an, long = RF digital-to-analog converter}
\DeclareAcronym{SDR}{short = SDR, short-indefinite = an, long = software-defined radio}
\DeclareAcronym{SoC}{short = SoC, short-indefinite = an, long = system on chip}

\NewDocumentCommand{\sbmathrm}{m}{\sb{\mathrm{#1}}}
\ExplSyntaxOn
\RenewDocumentCommand { \_ } {}
  { \mode_if_math:TF { \sbmathrm } { \textunderscore } }
\ExplSyntaxOff

\makeatletter
\NewDocumentCommand{\tablenotemark}{m}{\@fnsymbol{#1}}
\NewDocumentCommand{\tablenote}{m}{\textsuperscript{\tablenotemark{#1}}}
\NewDocumentCommand{\tablenotetext}{m m}{%
  \ifnumequal{#1}{1}{\par\vspace{\topsep}}{}%
  \par\tablenote{#1}#2%
}
\makeatother

\DeclareMathOperator{\sinc}{sinc}
\DeclareSIUnit{\dBm}{dBm}

\begin{document}

%
%
%
\title{Edge-Combining Subharmonic Mixer-First Receiver}

\author{\IEEEauthorblockN{Eric Danson and Jeffrey S. Walling}
        \IEEEauthorblockA{\textit{Bradley Dept.\ of Electrical and Computer Engineering} \\
        Virginia Tech, Blacksburg, VA, USA \\
        \{ecdanson, jswalling\}@vt.edu}
}


\maketitle%

%
%
\begin{abstract}
A \qtyrange{1}{10}{\giga\hertz} subharmonic mixer-first receiver using edge combining to drive N-path switches from 50\% duty-cycle clocks at half the \ac{LO} frequency avoids the gain penalty of harmonic mixing.
Inverter-based differential baseband amplifiers with back-gate feedback linearization set their gain by process-defined capacitance ratios, using no additional components and consuming no additional power.
A \qty{22}{\nano\meter} \ac{FD-SOI} prototype receiver achieves a conversion gain of \qtyrange{12.8}{18.1}{\decibel} and a noise figure of \qtyrange{5.7}{11}{\decibel} across the band.
It achieves up to \qty{15.4}{\dBm} out-of-band IIP3 while consuming \qtyrange{16.7}{22.3}{\milli\watt}.
\end{abstract}

\acresetall%


\acresetall%

%
%
%
\section{Introduction}

Integrated N-path mixer-first receivers can cover wide \ac{RF} bands and provide high-Q \ac{RF} filtering around a digitally-controlled center frequency for use in \acp{SDR}, but two bottlenecks limit how far this approach scales in frequency.
First, the switches require \(N \) non-overlapping clock phases with \(1/N \) duty cycle at the \ac{LO} frequency \(f\_{LO} \), usually derived from a master clock at a multiple of \(f\_{LO} \).
At multi-GHz \ac{LO} frequencies, the divider and switch drivers dominate the receiver power~\cite{purushothaman2020capstack,lin2014currentreuse}.
Driving the mixer at a subharmonic and downconverting with the \(n \)th \ac{LO} harmonic relaxes the clock frequency, but the conversion gain falls as \(\sinc\left(n/N \right) \), which also degrades noise figure~\cite{flete2022,rena2024}.
Second, the baseband amplifiers, rather than the switches, often limit the linearity and bandwidth of passive mixer-first receivers~\cite{krishnamurthy2021,razavi2022,lien2018}.

This work addresses both clock power and baseband linearity bottlenecks in mixer-first receivers.
Edge combining, previously used for frequency multiplication in \acp{RF-DAC}~\cite{nguyen2022}, is moved into the mixer switches: series-stacked switches driven by 50\%-duty-cycle phases at \(f\_{LO}/2 \) synthesizes the 25\% switching function directly at the RF input, so the receiver downconverts at the fundamental of the effective \ac{LO} without the harmonic-mixing loss.
The baseband uses inverter-based amplifiers with back-gate feedback linearization~\cite{danson2026}, whose gain is set by a well-controlled front-to-back gate capacitance ratio in \ac{FD-SOI} technology and costs no additional power nor components.

\Cref{sec:background} provides a review of edge combining.
\Cref{sec:architecture} describes the receiver architecture, followed in \cref{sec:results} by measurement results.
Finally, \cref{sec:conclusion} ends with conclusions.

\begin{figure}
  \centering
  \includegraphics[width = 0.9\columnwidth]{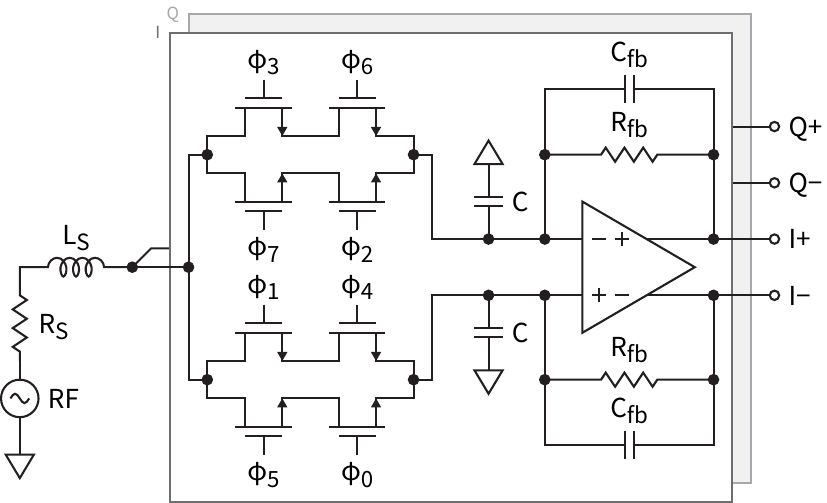}
  \caption{Proposed subharmonic mixer-first receiver architecture with edge combining for 2\(\times \) clock frequency multiplication.%
    \label{fig:n-path_mixer_edge_combining}}
\end{figure}

\section{Background%
  \label{sec:background}}

Edge combining is a frequency multiplication technique that works by interleaving multiple time-delayed versions of a lower-frequency clock signal with 50\% duty cycle.
The rising and falling edges of the delayed phases are combined by a logical AND operation, and the output signal switches at a multiple of the input frequency, determined by the number of phases.
Edge combining in N-path filters and mixers extends the N-path principle to the clock section and relaxes the \ac{LO} generation requirements.
N-path mixers, which multiply the \ac{RF} input by a square wave \ac{LO}, downconvert \ac{LO} harmonics with increasing loss.
For an ideal N-path mixer driven by non-overlapping clocks with \(1/N \) duty cycle, the magnitude of the \(n \)th harmonic voltage conversion gain from the \ac{RF} input at a frequency \(f\_{RF} = nf\_{LO} \) to a single baseband path is
\begin{equation}
  \left\lvert G_{\mathrm{v,}n} \right\rvert = \sinc\left(\frac{n}{N} \right)
\end{equation}
in normalized sinc notation.
For example, subharmonic downconversion with the second harmonic of a four-path mixer gives \(\left\lvert G\_{v,2} \right\rvert = \sinc\left(1/2 \right) \approx 0.64 \) (\qty{-3.9}{\decibel}), compared with \(\left\lvert G\_{v,2} \right\rvert = \sinc\left(1/4 \right) \approx 0.90 \) (\qty{-0.9}{\decibel}) at the fundamental.
Edge combining increases the fundamental \ac{LO} harmonic presented to the \ac{RF} input and avoids the lower conversion gain from using higher harmonics.

%
%
%
\section{Receiver Architecture%
  \label{sec:architecture}}

\subsection{Edge Combining}

\begin{figure}
  \centering
  \includegraphics[width = 0.6\columnwidth]{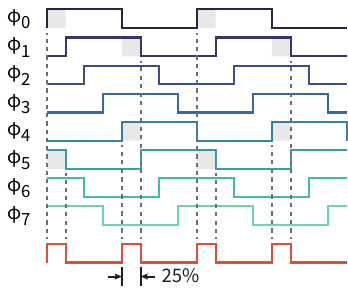}
  \caption{Edge combining waveforms.%
    \label{fig:edge_combining}}
\end{figure}

Each of the four baseband paths uses two parallel stacks of two series-connected switches in place of a single switch (\cref{fig:n-path_mixer_edge_combining}).
The clock phase separation in the series switches is \(\qty{180}{\degree} + \qty{360}{\degree}/\left(N\cdot M \right) = \qty{225}{\degree} \) where \(M = 2 \) is the clock frequency multiplication factor, and \(N\cdot M = 8 \) is the total number of clock phases.
A series pair conducts only when both gate phases are high, implementing the logical AND of two 50\%-duty-cycle phases taken from an eight-phase set at \(f\_{clk} = f\_{LO}/2 \).
The phase offset sets the overlap to \(T\_{LO}/4 \).
The parallel connection ORs two such pulses spaced by \(T\_{clk}/2 \), so each path conducts with a 25\% duty cycle at twice the clock frequency (\cref{fig:edge_combining}).
Because every clock net toggles at \(f\_{LO}/2 \) with 50\% duty cycle, the divider and buffers avoid generating narrow pulses at \(f\_{LO} \).
The trade-offs are that stacked switches need approximately twice the width as a single switch for the same on-resistance, and the extra stacks add parasitic capacitance at the \ac{RF} input node.
That capacitance limits higher frequency multiplication factors and the upper end of the \ac{RF} range.


\subsection{Clock Generation}

\begin{figure}
  \centering
  \includegraphics[width = 0.95\columnwidth]{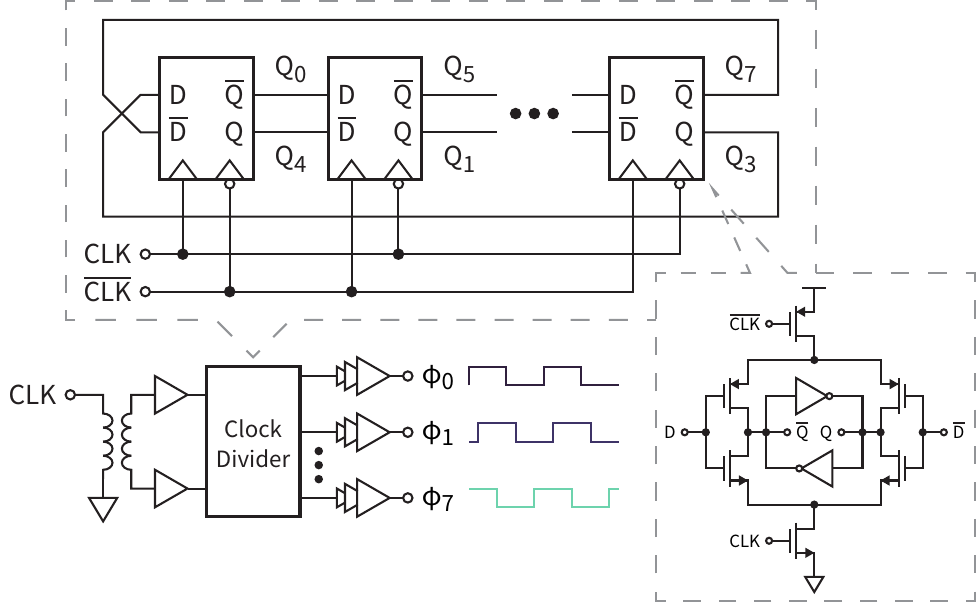}
  \caption{Multi-phase clock generation.%
    \label{fig:clock_generation}}
\end{figure}

An external single-ended sinusoidal clock signal at \(2f\_{LO} \) is converted to differential by an on-chip transformer balun, squared up by input buffers matched to \qty{50}{\ohm}, and divided by a four-stage Johnson counter into eight 50\%-duty-cycle phases (\cref{fig:clock_generation}).
Each phase drives an inverter chain with sizing optimized for jitter and power~\cite{overeem2023}.
Clock phases are distributed to the switches through an H-tree network for equal propagation delay, which is critical because phase skew directly changes the edge-combined pulse width and the I/Q balance.

\subsection{Baseband Amplifiers and Filters}

At baseband frequencies, the large loop gain available makes negative feedback a suitable linearization technique to reduce distortion, particularly if the feedback factor depends on a well-matched ratio in the semiconductor process rather than absolute values.
Feeding the transistor output to the back gate in \ac{FD-SOI} technology exploits the linearizing effect of negative feedback with a closed-loop voltage gain
\begin{equation}
  A\_{v} = \frac{A_{0}}{1 + \chi A_{0}} = -\frac{g\_{m}r\_{o}}{1 - \chi g\_{m}r\_{o}}
\end{equation}
where \(A_{0} = -g\_{m}r\_{o} \) is the intrinsic gain~\cite{danson2026}, and the feedback factor is the back-gate capacitive coupling ratio
\begin{equation}\label{eq:coupling_ratio}
  \chi = -\frac{g\_{mb}}{g\_{m}} = -\frac{C\_{box}\parallel C\_{Si}}{C\_{ox}}
\end{equation}
where \(C\_{box} \), \(C\_{Si} \), and \(C\_{ox} \) are the buried oxide, thin silicon channel, and gate oxide capacitances, respectively~\cite{agarwal2020}.
Considering the intrinsic gain as a first-order system
\begin{equation}
  A\_{0}\left(s \right) = -\frac{g\_{m}r\_{o}}{1 + s/\omega\_{p}}
\end{equation}
with a dominant pole \(\omega\_{p} \), the root of the characteristic equation
\begin{equation}
  \begin{aligned}
    1 + \chi A\_{0}\left(s \right) &= 0 \\
    s &= -\left(1 + g\_{mb}r\_{o} \right)\omega\_{p}
  \end{aligned}
\end{equation}
is real and negative.
The closed-loop pole is in the left-half plane, so the system is stable with back-gate feedback.
When the loop gain magnitude \(\left|\chi g\_{m}r\_{o} \right| \gg 1 \), the closed-loop gain is approximately \(1/\chi = -g\_{m}/g\_{mb} \) and is set by the transistor's front-to-back gate oxide capacitance ratio in the semiconductor process.

\begin{figure}
  \centering
  \includegraphics[width = 0.6\columnwidth]{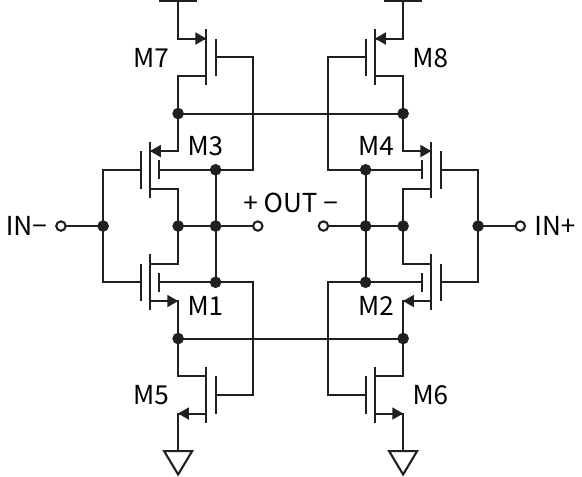}
  \caption{Dual common-mode feedback differential amplifier with back-gate feedback linearization~\cite{danson2026}.%
    \label{fig:ccs_backgate_diff_dcmfb}}
\end{figure}

Back-gate feedback makes the baseband amplifier gain approximately independent of the load resistance.
The small-signal voltage gain of the complementary common-source amplifier with back-gate feedback
\begin{equation}
  A\_{v} = -\frac{\left(g\_{mn} + g\_{mp} \right)\left(r\_{on}\parallel r\_{op} \right)}{1 + \left(g\_{mbn} + g\_{mbp} \right)\left(r\_{on}\parallel r\_{op} \right)} \approx -\frac{g\_{mn} + g\_{mp}}{g\_{mbn} + g\_{mbp}}
\end{equation}
shows the same intrinsic gain \(-\left(g\_{mn} + g\_{mp} \right)\left(r\_{on}\parallel r\_{op} \right) \) through the front gate as in open loop (\cref{fig:ccs_backgate_diff_dcmfb}).
Adding feedback reduces the output resistance \(r\_{on}\parallel r\_{op} \) by the loop gain \(\left(g\_{mbn} + g\_{mbp} \right)\left(r\_{on}\parallel r\_{op} \right) \), so the overall gain is approximately the front- and back-gate conductance ratio \(1/\chi \) from \cref{eq:coupling_ratio} in saturation while requiring no additional components and consuming no additional power compared to without feedback~\cite{danson2026}.

%
%
\section{Measurement Results%
  \label{sec:results}}

\begin{figure}
  \centering
  \includegraphics[width = 0.7\columnwidth]{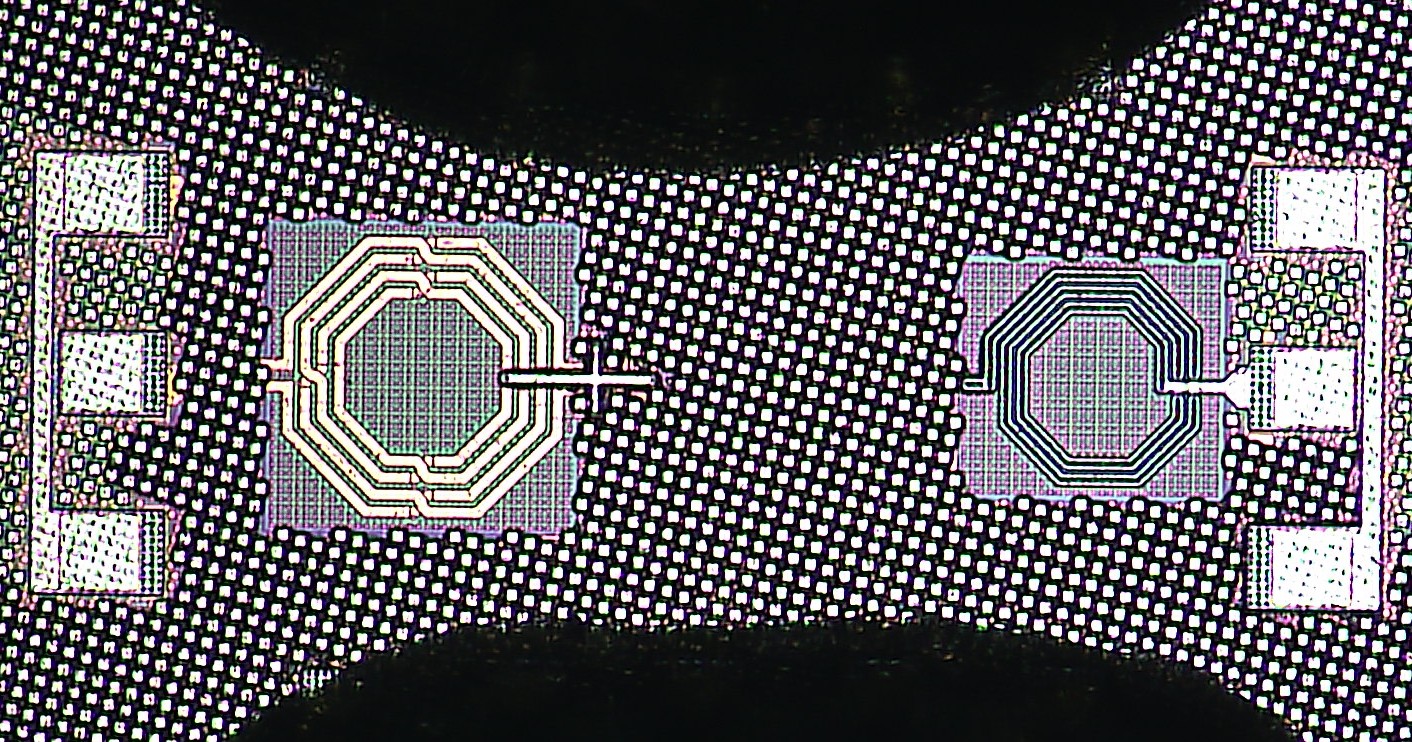}
  \caption{Die micrograph of the \qty{22}{\nano\meter} \ac{FD-SOI} prototype receiver front-end.%
    \label{fig:die_micrograph}}
\end{figure}

A prototype mixer-first receiver front-end was implemented in GlobalFoundries' \qty{22}{\nano\meter} \ac{FD-SOI} technology using \qty{0.8}{\volt} thin-oxide devices for the clock generation circuitry and the core switches, with \qty{1.8}{\volt} thick-oxide I/O devices for the baseband amplifiers to extend the linear output operating range (\cref{fig:die_micrograph}).
Wire bonding pads on the top and bottom (not visible) supply power and connect the differential baseband I/Q outputs to external buffers capable of driving a \qty{50}{\ohm} load.
The test circuit occupies a total area of \qty{0.335}{\milli\meter\squared} including pads.
The active or functional block area is \qty{0.0979}{\milli\meter\squared} when including inductors and \qty{0.0345}{\milli\meter\squared} without inductors.
It is noted that one of the inductors is used to receive the clock from off chip and would be obviated in \iac{SoC} implementation.
A linear fit \(y = mx + b \) of the clock power consumption has a slope \(m = \qty{0.674}{\milli\watt\per\giga\hertz} \) and an intercept \(b = \qty{5.28}{\milli\watt} \).
The mean baseband power is \qty{10.0}{\milli\watt}, and the total power consumption ranges from \qty{16.7}{\milli\watt} at \qty{1}{\giga\hertz} to \qty{22.3}{\milli\watt} at \qty{10}{\giga\hertz}.

\begin{figure}
  \centering
  \includegraphics[width = 0.9\columnwidth]{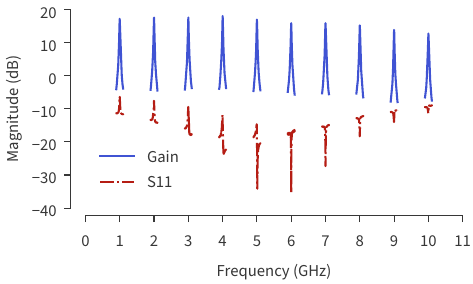}
  \caption{Measured conversion gain and $S_{11}$.%
    \label{fig:conversion_gain_s11_sweep}}
\end{figure}

\begin{figure}
  \centering
  \includegraphics[width = 0.9\columnwidth]{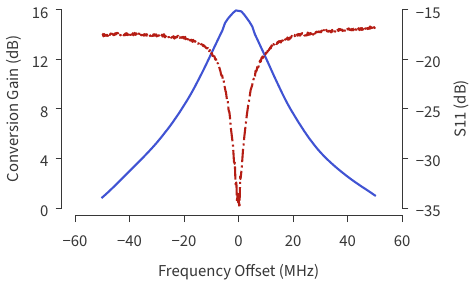}
  \caption{Measured conversion gain and S11 centered at \(f\_{LO} = \qty{6}{\giga\hertz} \).%
    \label{fig:conversion_gain_s11_6G}}
\end{figure}

\begin{figure}
  \centering
  \includegraphics[width = 0.9\columnwidth]{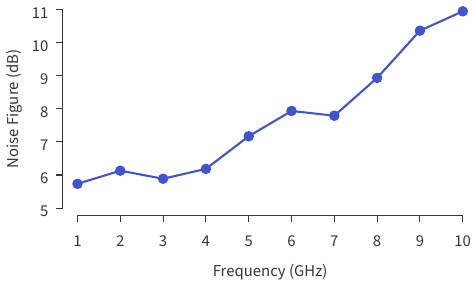}
  \caption{Measured noise figure.%
    \label{fig:noise_figure_sweep}}
\end{figure}

The conversion gain ranges from \qty{12.8}{\decibel} at \qty{10}{\giga\hertz} to \qty{18.1}{\decibel} at \qty{4}{\giga\hertz} (\cref{fig:conversion_gain_s11_sweep}).
The baseband \qty{-3}{\decibel} bandwidth is \qty{10}{\mega\hertz} on either side of the \ac{LO} frequency for a \qty{20}{\mega\hertz} total channel bandwidth (\cref{fig:conversion_gain_s11_6G}).
Input matching is quantified by S-parameters.
\(S_{11} \) reaches a minimum of \qty{-35}{\decibel} at \qty{6}{\giga\hertz} because of the series \ac{RF} input inductor used to improve impedance matching, and \(S_{11} \) increases towards the upper and lower bounds of the operating frequency range.
The noise figure was measured using the Y-factor method and ranges from \qty{5.73}{\decibel} at \qty{1}{\giga\hertz} to \qty{11.0}{\decibel} at \qty{10}{\giga\hertz} (\cref{fig:noise_figure_sweep}).

\begin{figure}
  \centering
  \includegraphics[width = 0.9\columnwidth]{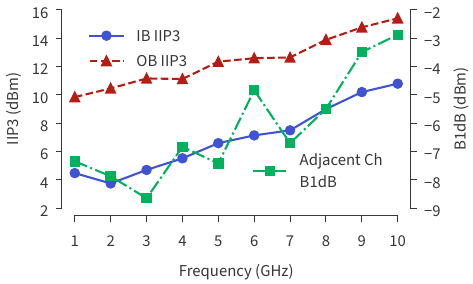}
  \caption{Measured in-band IIP3, out-of-band IIP3, and adjacent channel B1dB with a blocker offset \(\Delta f / f\_{-3dB} = 1 \).%
    \label{fig:iip3_b1db_sweep}}
\end{figure}

\begin{figure}
  \centering
  \includegraphics[width = 0.9\columnwidth]{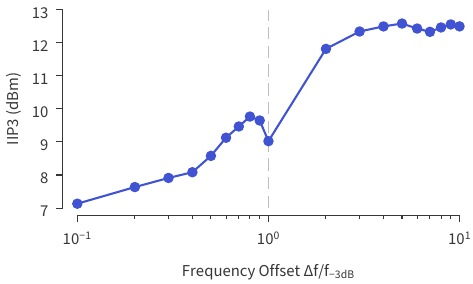}
  \caption{Measured IIP3 centered at \(f\_{LO} = \qty{6}{\giga\hertz} \).%
    \label{fig:iip3_sweep_offset_6G}}
\end{figure}

Linearity is characterized by the \ac{IIP3} and was measured with two tones \(f_{1} = f\_{LO} + \Delta f + f\_{IM3} \) and \(f_{2} = f\_{LO} + 2\Delta f + f\_{IM3} \).
The \ac{IM3} products have a constant baseband frequency \(\left(2f_{1} - f_{2} \right) - f\_{LO} = f\_{IM3} = \qty{0.5}{\mega\hertz} \).
\Ac{IIP3} increases with offset frequency \(\Delta f \), from \qtyrange{3.74}{10.8}{\decibel} in-band to \qtyrange{9.84}{15.4}{\decibel} out-of-band over the full operating range (\cref{fig:iip3_b1db_sweep,fig:iip3_sweep_offset_6G}).

\begin{table*}
  \centering
  \caption{Performance Comparison%
    \label{tab:performance_comparison}}
  \begin{booktabs}{ccccc}
    \toprule
    & \textbf{This Work} & Zijlma et al.~\cite{zijlma2025} & Huang and Molnar~\cite{huang2021} & Lien et al.~\cite{lien2018} \\ \midrule
    Technology & \qty{22}{\nano\meter} FD-SOI & \qty{22}{\nano\meter} FD-SOI & \qty{45}{\nano\meter} SOI & \qty{45}{\nano\meter} SOI \\
    RF Input & Single-ended & Single-ended & Single-ended & Differential \\
    Frequency (\unit{\giga\hertz}) & 1--10 & 1--10 & 3.7--6.5 & 0.2--8 \\
    BB BW (\unit{\mega\hertz}) & 10 & 10 & --- & 10 \\
    Gain (\unit{\decibel}) & 12.8--18.1 & 10--13 & --- & 21 \\
    NF (\unit{\decibel}) & 5.7--11 & 4.7--7.0 & 2.4--4.7 & 2.3--7.1 \\
    IB IIP3 (\unit{\dBm}) & 3.7--10.8 & 1--5 & -32\tablenote{1} & 0\tablenote{1} \\
    OOB IIP3 (\unit{\dBm}) (@\(\Delta f/f\_{-3dB} \)) & 9.8--15.4 (10) & 21--23 (50) & 28 & 39 (8) \\
    Supply (\unit{\volt}) & 0.8, 1.8 & 0.9 & --- & 1.2 \\
    Static Power (\unit{\milli\watt}) & 15.3 & 0 & 36 & 50 \\
    Dynamic Power (\unit{\milli\watt\per\giga\hertz}) & 0.671 & 3.1 & 16.4 & 30 \\
    Active Area (\unit{\milli\meter\squared}) & 0.1 & 0.05 & --- & 0.8 \\ \bottomrule
  \end{booktabs}
  \tablenotetext{1}{Estimated from plots}
\end{table*}

%
%
\section{Conclusion%
  \label{sec:conclusion}}

A subharmonic mixer-first receiver in \qty{22}{\nano\meter} \ac{FD-SOI} moves edge combining into the N-path switches.
The clock path runs at half the \ac{LO} frequency with 50\% duty cycle, while the \ac{RF} input sees a fundamental 25\% switching function, so the receiver avoids the conversion gain and noise penalty of harmonic mixing.
Back-gate-feedback inverter amplifiers provide a process-ratio-defined baseband gain at no extra power. The prototype operates from \qtyrange{1}{10}{\giga\hertz} with \qtyrange{12.8}{18.1}{\decibel} conversion gain, \qtyrange{5.7}{11}{\decibel} noise figure, and \qtyrange{16.7}{22.3}{\milli\watt} total power in \qty{0.0345}{\milli\meter\squared} of active area excluding inductors.

%
%
\section*{Acknowledgment}

We would like to acknowledge GlobalFoundries' University Partnership Program for providing process development kits.

%
%
\balance%
\bibliography{bibliography/IEEEabrv, bibliography/IEEEbstctl, bibliography/references}

@STRING{IEEE_J_JSSC       = "{IEEE} J. Solid-State Circuits"}

@IEEEtranBSTCTL{IEEEexample:BSTcontrol,
  CTLuse_forced_etal = "yes",
  CTLmax_names_forced_etal = "6"
}

@inproceedings{agarwal2020,
  title = {{{BSIM-IMG}}: {{Advanced Model}} for {{FDSOI Transistors}} with {{Back Channel Inversion}}},
  shorttitle = {{{BSIM-IMG}}},
  booktitle = {2020 4th {{IEEE Electron Devices Technology}} \& {{Manufacturing Conference}} ({{EDTM}})},
  author = {Agarwal, H. and Kushwaha, P. and Dasgupta, A. and {Y-Kao}, M. and Morshed, T. and Workman, G. and Shanbhag, K. and Li, X. and Vinothkumar, V. and Chauhan, Y. S. and Salahuddin, S. and Hu, C.},
  year = {2020},
  month = apr,
  pages = {1--4},
  doi = {10.1109/EDTM47692.2020.9117979},
  urldate = {2025-09-23}
}

@inproceedings{danson2026,
  title = {Inverter-{{Based Differential Amplifiers With Back-Gate Feedback Linearization}}},
  booktitle = {2026 24th {{IEEE Interregional NEWCAS Conference}} ({{NEWCAS}})},
  author = {Danson, Eric and Walling, Jeffrey},
  year = 2026,
  month = jun,
  pages = {1--4},
  issn = {2474-9672},
  doi = {10.1109/NewCAS64543.2026.11674000},
  urldate = {2026-11-06}
}

@inproceedings{flete2022,
  title = {A \texttimes 3 {{Sub-Sampling Mixer}} for a 77 {{GHz Automotive Radar Receiver}} in 28 Nm {{FD-SOI CMOS Technology}}},
  booktitle = {2022 {{IEEE}} 22nd {{Topical Meeting}} on {{Silicon Monolithic Integrated Circuits}} in {{RF Systems}} ({{SiRF}})},
  author = {Flete, Alexandre and Viallon, Christophe and Cathelin, Philippe and Parra, Thierry},
  year = 2022,
  month = jan,
  pages = {52--54},
  issn = {2474-9761},
  doi = {10.1109/SiRF53094.2022.9720061},
  urldate = {2026-01-12}
}

@inproceedings{huang2021,
  title = {A 3.7-6.{{5GHz}} 8-{{Phase N-Path Mixer-First Receiver}} with {{LO Overlap Suppression Achieving}} {$<$}{{5dB NF}} and {$>$}{{5dBm OOB B1dB}}},
  booktitle = {2021 {{IEEE Radio Frequency Integrated Circuits Symposium}} ({{RFIC}})},
  author = {Huang, Shimin and Molnar, Alyosha},
  year = 2021,
  month = jun,
  pages = {87--90},
  issn = {2375-0995},
  doi = {10.1109/RFIC51843.2021.9490451},
  urldate = {2025-09-08}
}

@article{lien2018,
  title = {Enhanced-{{Selectivity High-Linearity Low-Noise Mixer-First Receiver With Complex Pole Pair Due}} to {{Capacitive Positive Feedback}}},
  author = {Lien, Yuan-Ching and Klumperink, Eric A. M. and Tenbroek, Bernard and Strange, Jon and Nauta, Bram},
  year = 2018,
  month = may,
  journal = IEEE_J_JSSC,
  volume = {53},
  number = {5},
  pages = {1348--1360},
  issn = {1558-173X},
  doi = {10.1109/JSSC.2018.2791490},
  urldate = {2026-11-11}
}

@article{nguyen2022,
  title = {A Mm-{{Wave Switched-Capacitor RFDAC}}},
  author = {Nguyen, Hieu Minh and Walling, Jeffrey Sean and Zhu, Anding and Staszewski, Robert Bogdan},
  year = 2022,
  month = apr,
  journal = IEEE_J_JSSC,
  volume = {57},
  number = {4},
  pages = {1224--1238},
  issn = {1558-173X},
  doi = {10.1109/JSSC.2022.3142718},
  urldate = {2026-09-02}
}

@inproceedings{overeem2023,
  title = {Inverter {{Chain Buffer Optimization}} for {{N}}-Path {{Filter Switch Drivers}} and {{Validation}} through {{Simulations}} in 22nm {{FD-SOI Technology}}},
  booktitle = {2023 {{IEEE International Symposium}} on {{Circuits}} and {{Systems}} ({{ISCAS}})},
  author = {Overeem, Wouter T. and Alink, Mark S. Oude and Nauta, Bram},
  year = 2023,
  month = may,
  pages = {1--5},
  issn = {2158-1525},
  doi = {10.1109/ISCAS46773.2023.10181574},
  urldate = {2025-09-08}
}

@article{rena2024,
  title = {A 0.4--1.8-{{GHz Quarter-Rate Subsampling Mixer-First Direct Down-Conversion RF Front-End}}},
  author = {Rena, Rakesh Varma and Kammari, Raviteja and Pasupureddi, Vijay Shankar},
  year = 2024,
  month = mar,
  journal = {IEEE Transactions on Very Large Scale Integration (VLSI) Systems},
  volume = {32},
  number = {3},
  pages = {552--563},
  issn = {1557-9999},
  doi = {10.1109/TVLSI.2023.3342022},
  urldate = {2026-01-12}
}

@article{zijlma2025,
  title = {Analysis and {{Design}} of a {{Low-Loss}} 1--10 {{GHz Capacitive Stacking N-Path Filter}}/{{Mixer}}},
  author = {Zijlma, Emiel and van Zanten, Stef and Plompen, Roel and Klumperink, Eric A. M. and van der Zee, Ronan A. R. and Nauta, Bram},
  year = 2025,
  month = feb,
  journal = IEEE_J_JSSC,
  volume = {60},
  number = {2},
  pages = {367--381},
  issn = {1558-173X},
  doi = {10.1109/JSSC.2024.3407241},
  urldate = {2026-07-30}
}

@ARTICLE{purushothaman2020capstack,
  author={Purushothaman, Vijaya Kumar and Klumperink, Eric A. M. and Clavera, Berta Trullas and Nauta, Bram},
  journal=IEEE_J_JSSC, 
  title={A Fully Passive RF Front End With 13-dB Gain Exploiting Implicit Capacitive Stacking in a Bottom-Plate N-Path Filter/Mixer}, 
  year={2020},
  volume={55},
  number={5},
  pages={1139-1150},
  doi={10.1109/JSSC.2019.2959489}}

@ARTICLE{lin2014currentreuse,
  author={Lin, Fujian and Mak, Pui-In and Martins, Rui P.},
  journal=IEEE_J_JSSC, 
  title={An RF-to-BB-Current-Reuse Wideband Receiver With Parallel N-Path Active/Passive Mixers and a Single-MOS Pole-Zero LPF}, 
  year={2014},
  volume={49},
  number={11},
  pages={2547-2559},
  doi={10.1109/JSSC.2014.2354647}}

@ARTICLE{krishnamurthy2021,
  author={Krishnamurthy, Sashank and Maksimovic, Filip and Iotti, Lorenzo and Niknejad, Ali M.},
  journal={IEEE Transactions on Microwave Theory and Techniques}, 
  title={Analysis and Design of Submilliwatt Interference-Tolerant Receivers Leveraging N-Path Filter-Based Translational Positive Feedback}, 
  year={2021},
  volume={69},
  number={7},
  pages={3496-3509},
  doi={10.1109/TMTT.2021.3076823}}

@ARTICLE{razavi2022,
  author={Razavi, Hossein and Razavi, Behzad},
  journal=IEEE_J_JSSC, 
  title={A 0.4–6 GHz Receiver for Cellular and WiFi Applications}, 
  year={2022},
  volume={57},
  number={9},
  pages={2640-2657},
  doi={10.1109/JSSC.2022.3146080}}

\end{document}